# Magnetic Multipoles in a Ruthenate $Ca_3Ru_2O_7$

S. W. Lovesey[1,2], D. D. Khalyavin[1], and G. van der Laan[2]

[1] ISIS Facility, STFC, Didcot, Oxfordshire OX11 0QX, UK

[2] Diamond Light Source, Didcot, Oxfordshire OX11 0DE, UK

**Abstract** Compulsory Dirac multipoles in the bilayer perovskite $Ca_3Ru_2O_7$ are absent in published analyses of experimental data. In a first step at correcting knowledge of the magnetic structure, we have analysed existing Bragg diffraction patterns gathered on samples held well below the Néel temperature at which A-type antiferromagnetic order of axial dipoles spontaneously develops. Patterns were gathered with neutrons, and linearly polarized x-rays tuned in energy to a ruthenium atomic resonance. Neutron diffraction data contains solid evidence of Dirac dipoles (anapoles or toroidal moments). No such conclusion is reached with existing x-ray diffraction data, which instead is ambiguous on the question. To address this shortcoming by future experiments, we calculated additional diffraction patterns. Chiral order of Dirac multipoles is allowed by magnetic space-group $P_Cna2_1$, and it can be exposed in Bragg diffraction using circularly polarized x-rays. Likewise, a similar experiment can expose a chiral order of axial dipoles. A magnetic field applied parallel to the b-axis creates a ferrimagnetic structure in which bulk magnetization arises from field-induced nonequivalent Ru sites (magnetic space-group $Pm'c'2_1$).

## I. INTRODUCTION

The laminar perovskite $Ca_3Ru_2O_7$ contains one ruthenium-oxygen bilayer per formula unit and crystallizes in an orthorhombic space group with a unit cell containing four formula units [1, 2]. While the ruthenate is structurally similar to a bilayer cuprate, e.g., YBCO ($YBa_2Cu_3O_{6+x}$), not all the strange physics of underdoped cuprates is present [3-6]. It is a low carrier-concentration material on the verge of a metal-insulator transition, with a colossal magneto-resistance [7, 8]. At a Néel temperature $T_N \approx 56$ K the material is metallic. On reaching 48 K the resistivity along the c-axis abruptly increases and continues to increase with cooling. Resistivity in the a-b plane is not changed so dramatically and becomes only weakly dependent on temperature below 48 K, with evidence of a quasi-two-dimensional metallic ground state below 30 K [2, 9].

Lattice constants of $Ca_3Ru_2O_7$ jump at the first-order metal-to-nonmetal phase transition at 48 K without a change of the polar space-group symmetry $Bb2_1m$ illustrated in Fig. 1 [1]. Magnetic reflections present below 48 K in neutron and resonant x-ray Bragg diffraction patterns are consistent with A-type antiferromagnetic order of axial dipoles and a propagation vector along the c-axis [1, 10 11]. The $RuO_2$ bilayers depicted in Fig. 1 (after reference [1]) are ferromagnetically ordered in the a-b plane and are antiferromagnetically stacked normal to the plane.

Dirac multipoles are a compulsory component of a magnetic structure when a centre of inversion symmetry is absent in sites occupied by magnetic ions. To date, these multipoles have been omitted from all experimental studies of $Ca_3Ru_2O_7$, e.g., Zhu *et al*. and Sokolov *et al*. [12, 13]. We address this shortcoming with an analysis of published neutron and resonant x-ray Bragg diffraction patterns [10, 11], taking account of Dirac multipoles and axial (parity-even) multipoles [14]. For Dirac dipoles, also called anapoles or toroidal dipoles, a persuasive case can be extracted from a neutron diffraction pattern at hand [10]. Resonant x-ray Bragg diffraction data are inconclusive, however.

By way of a potential remedy, we propose additional experiments that can vastly improve the knowledge of multipoles in the ruthenate in question. Specifically, we predict that a handed setting will reveal chiral order among the primary and secondary axial dipoles. Likewise, chiral order exists within Dirac multipoles of even rank, e.g., the monopole and quadrupoles. Circular polarization in the primary x-ray beam, and collinear Bragg and propagation vectors, furnishes a suitable handed setting. A magnetic field applied parallel to the b-axis creates a ferrimagnetic structure in which bulk magnetization arises from field-induced nonequivalent Ru sites.

The magnetic structure of $Ca_3Ru_2O_7$ precludes a direct observation of Dirac multipoles by neutron diffraction accomplished for $SmAl_2$ that presents a diamond-type structure [15]. Set against controversial evidence for Dirac multipoles in the pseudo-gap phase of YBCO - the very existence of magnetic Bragg spots in neutron diffraction patterns is an ongoing debate [16-19] - our evidence in favour of anapoles in $Ca_3Ru_2O_7$ is compelling.

## II. CRYSTAL AND MAGNETIC STRUCTURES

The parent structure of $Ca_3Ru_2O_7$ is orthorhombic $Bb2_1m$ (#36, polar crystal-class mm2 ($C_{2v}$), cell lengths a ≈ 5.3677, b ≈ 5.5356, c ≈ 19.5219 Å at 8 K [1]) that is a B-face centred cell with glide symmetry along the b-axis, and the structure is more distorted than the companion double-layer ruthenate $Sr_3Ru_2O_7$ (Bbcb, #68). Below 48 K, ruthenium axial dipoles align principally along the b-axis derived from magnetic space-group $P_Cna2_1$ (#33.154 [20], crystal class mm21′). Crystal and magnetic structures are displayed in Fig. 1. Ruthenium ions occupy sites 8b with no symmetry and coordinates (1/4, 0.4008, 1/4). Note that general coordinates (x, y, z) are not fixed by symmetry in a polar crystal, and our use of x = z = 1/4 is an approximation, albeit an excellent approximation [1]. A basis (1, 0, 0), (0, 0, 1), (0, −1, 0) with respect to the $Bb2_1m$ paramagnetic space-group defines orthonormal principal axes ($\xi$, $\eta$, $\zeta$), with Miller indices $h = H_o$, $k = L_o$, $l = -K_o$. Magnetic Bragg spots are indexed by Miller indices $(h + k) = (H_o + L_o)$ odd, a consequence of anti-translation in the space-group, and $(H_o + K_o)$ even. Looking ahead, Section IV. C is devoted to the influence of a magnetic field that creates ferrimagnetic order. With the field applied along the b-axis, purely magnetic reflections occur for $H_o = 0$, $K_o$ odd and all $L_o$.

Electronic degrees of freedom present in the ground state of a Ru ion are encapsulated in multipoles $\langle U^K_Q \rangle$ with rank K and projections Q that obey $-K \leq Q \leq K$. Angular brackets $\langle ... \rangle$ denote the expectation value of the enclosed spherical tensor operator, defined for neutron and x-ray diffraction in references [21, 22]. The operator $U^K_Q$ is time-odd, of course, and its parity is labelled $\sigma_\pi = +1 (-1)$ axial (polar). Hermitian operators yield multipoles that obey $\langle U^K_Q \rangle^* = (-1)^Q \langle U^K_{-Q} \rangle$, and $\langle U^K_Q \rangle = \langle U^K_Q \rangle' + i\langle U^K_Q \rangle''$ is our convention for real and imaginary parts. Multipoles are evaluated in principal axes, e.g., the dipole $\langle U^1_\eta \rangle$ is parallel to the crystal c-axis $\equiv$ propagation vector (Cartesian and spherical components of a vector **R** are related by $R_\xi = (R_{-1} - R_{+1})/\sqrt{2}$, $R_\eta = i(R_{-1} + R_{+1})/\sqrt{2}$ and $R_\zeta = R_0$).

Multipoles that are both parity-odd ($\sigma_\pi = -1$) and time-odd are referred to as Dirac multipoles. Dirac dipoles (anapoles) observed in magnetic neutron diffraction are simple products of spin **S** or orbital angular momentum **L** with the electronic position operator **n** [21, 22]. For these anapoles we use $\langle \mathbf{\Omega} \rangle_S = \langle \mathbf{S} \times \mathbf{n} \rangle$ and $\langle \mathbf{\Omega} \rangle_L = \langle \mathbf{L} \times \mathbf{n} - \mathbf{n} \times \mathbf{L} \rangle$, and it is noted that operators **S** and **n** commute whereas **L** and **n** do not commute. A monopole $\langle \mathbf{S} \cdot \mathbf{n} \rangle$ is allowed in resonant x-ray diffraction [23] but it is forbidden in neutron diffraction [22], while $\langle \mathbf{L} \cdot \mathbf{n} \rangle = 0$.

Unit-cell structure factors for Bragg diffraction are derived from $\Psi^K_Q = \{\sum \exp(i\mathbf{d} \cdot \mathbf{k}) \langle U^K_Q \rangle_\mathbf{d}\}$, where **k** is the Bragg wavevector. Sites 8b, with positions **d** in a cell, used by $Ru^{4+}$ ions are asymmetric. Symmetry operations in $P_Cna2_1$ lead to the central result,

$$\Psi^K_Q = \exp[i\pi(h+l)/2] \, [\exp(i\varphi) + (-1)^{h+l+Q} \exp(-i\varphi)] \quad (1)$$

$$\times \, [\langle U^K_Q \rangle - \sigma_\pi (-1)^{h+l+K} \langle U^K_{-Q} \rangle],$$

with a spatial phase angle $\varphi = (2\pi y L_o)$ and $y \approx 0.4008$ [1]. Site symmetry adds nothing to symmetry displayed by $\Psi^K_Q$, in contrast to many materials where site symmetry is hugely influential [24]. The leading phase factor in (1) for $\Psi^K_Q$ can be set aside in the calculation of Bragg intensities, for they depend on the absolute value of the unit-cell structure factor derived from $\Psi^K_Q$.

The energy levels the free $Ru^{4+}$ $4d^4$ ion in intermediate coupling are determined by the 4d-4d Coulomb interaction (with atomic Slater integrals $F_2 = 9.214$ eV and $F_4 = 6.095$ eV) and spin-orbit interaction (with parameter $\zeta_d = 160$ meV $= 1290$ cm$^{-1}$). The ground state is $^5D$ (S = 2 and L = 2 with cancelling opposite moments) which has a purity of 96%. The remaining 4% is made up of other SL terms which are mixed in the ground state by the spin-orbit interaction. According to the third Hund's rule the lowest energy level has total angular momentum J = 0, for which all magnetic multipoles are zero. To this atomic scenario we add the effect of the spin-orbit interaction and a spatially symmetric ligand field, which is a useful, albeit approximate, rendition of Ru sites that have no symmetry.

In pure octahedral symmetry, the $^5D$ splits into $^5E_g + ^5T_{1g}$ states. In weak ligand field the lowest energy state is the high-spin $t_{2g}^3 e_g^1$ ($^5E_g$) (see, e.g., [25]). Under spin-orbit interaction, the $^5$E splits into $\Gamma_1, \Gamma_2, \Gamma_3, \Gamma_4, \Gamma_5$ levels (in Koster notation, or $A_1, A_2, E, T_1, T_2$ in Schönflies notation), where the $\Gamma_1$ level has the lowest energy. A strong ligand field (typical cubic crystal-field parameter 10Dq ≈ 3 eV, appropriate for ions in the palladium group) results in a $t_{2g}^4$ ($^3T_{1g}$) ground state, which is low spin (S = 1). Under spin-orbit interaction, the $^3T_1$ splits into $\Gamma_1, \Gamma_4, \Gamma_3, \Gamma_5$ levels with multiplicities of 1, 3, 2, 3, respectively. These levels have an energy separation in the order of $\zeta_d$. Thus in all cases the total angular momentum in the ground state has the total symmetric representation $\Gamma_1$.

However, the magnetic exchange interaction $H_{ex}$ mixes higher J levels into the ground state if $H_{ex}$ is in the order of the spin-orbit interaction. This effect is clearly seen for Cr $d^4$ $^5D_0$ (with $\zeta_d$ ≈ 30 meV), which has no dichroism in the absence of J mixing [26]. For Ru $d^4$, where the spin-orbit interaction is much larger, the J mixing due to $H_{ex}$ is much smaller but still observable. J mixing does also occur due to hybridization with the neighbouring ligands, because the 4d bandwidth of several eV is much larger than $\zeta_d$.

### III. NEUTRON DIFFRACTION

Bao *et al*. [10] report Bragg intensities for a sample temperature 3.5 K indexed by Miller indices $h$ and $l$ even. In consequence, the spatial phase-factor in $\Psi^K_Q$ is {i sin($\varphi$)} for Q odd or cos($\varphi$) for Q even.

Dipoles (K = 1) possess either Q = 0 or Q = ±1. Thus, axial dipoles are aligned with the $\zeta$-axis (Q = 0) or the $\eta$-axis, according to the structure factor (1). Magnetic moments $\mu_\zeta = \langle 2S_\zeta + L_\zeta \rangle$ aligned with the $\zeta$-axis form a primary dipole-order. Secondary dipole-order arises from $\mu_\xi$ and $\mu_\eta$. Use of general coordinates x = z = 1/4 for Ru sites 8b eliminates the axial $\xi$-dipole from the structure factor. Axial dipoles along $\eta$- and $\zeta$-axes are 90° out of phase in the structure factor. In consequence, $\mu_\eta$ can only add intensity to Bragg spots with $h$ and $l$ different from zero, while there is no contribution to (0, $k$, 0) Bragg spots from $\mu_\eta$. The unit-cell structure factor confronted with Bragg spots contains quadrupoles (K = 2) and octupoles (K = 3) with even projections and a spatial phase cos($\varphi$). Values for the axial multipoles are inferred from experimental data [10].

Anapoles allowed in neutron diffraction include t = [3$\langle \Omega_\xi \rangle_S$ ($h_1$) − $\langle \Omega_\xi \rangle_L$ ($j_0$)] accompanied by a spatial phase sin($\varphi$). Radial integrals ($h_1$) and ($j_0$) in t are displayed in Fig. 2a using a dimensionless variable w = 12$\pi a_o$s, where $a_o$ is the Bohr radius and s = sin($\theta$)/$\lambda$ (atomic code due to R. D. Cowan [27]). Values of t are inferred from fitting our analytic expressions (2) and (3) for intensities to observed intensities. An anapole t′ = {$\langle n_\xi \rangle$ ($g_1$)} is also accompanied by a spatial phase sin($\varphi$) and its contribution to the unit-cell structure factor is 90° out of phase with t. We have tested our analytic expressions against experimental data for both t ≠ 0, t′ = 0 and, also, t = 0, t′ ≠ 0. In the latter case, inferred values of t′ are unrelated to the radial integral ($g_1$) displayed in Fig. 2a. We achieve better success with t versus a linear combination of ($h_1$) and ($j_0$) shown in Fig. 3.

Bragg spot intensities confronted with data are calculated from,

$$\Im(t) = [\{\mathbf{Q} \cdot \mathbf{Q}^*\} - |\mathbf{\kappa} \cdot \mathbf{Q}|^2], \qquad (2)$$

with the following amplitude **Q** derived from (1) and expressions recorded in reference [22],

$$Q_\xi \approx \kappa_\xi \kappa_\zeta \cos(\varphi) \langle j_2 \rangle [-q - p + p'],$$

$$Q_\eta \approx \kappa_\eta \kappa_\zeta \cos(\varphi) \langle j_2 \rangle [q - p - p'] - \sin(\varphi) \, t \, \kappa_\zeta, \qquad (3)$$

$$Q_\zeta \approx \cos(\varphi) [\langle j_0 \rangle \mu_\zeta + \langle j_2 \rangle \{\langle L_\zeta \rangle + (\kappa_\xi^2 - \kappa_\eta^2)(q + p'/2) + (3\kappa_\zeta^2 - 1)(p/2)\}] + \sin(\varphi) \, t \, \kappa_\eta,$$

$$\mathbf{\kappa} \cdot \mathbf{Q} \approx \kappa_\zeta \cos(\varphi) [\langle j_0 \rangle \mu_\zeta + \langle j_2 \rangle \{\langle L_\zeta \rangle + (\kappa_\xi^2 - \kappa_\eta^2)(3p'/2) + (5\kappa_\zeta^2 - 3)(p/2)\}].$$

A standard dipole approximation to the neutron scattering amplitude is obtained from (3) on setting $t = p = q = p' = 0$. Evidently, this approximation depends on $\{\langle j_0 \rangle \mu_\zeta\}$ and $\{\langle j_2 \rangle \langle L_\zeta \rangle\}$ arising from $Q_\zeta$, with $Q_\xi = Q_\eta = 0$. We go beyond this simple expression for the scattering amplitude through addition of some other multipoles allowed by magnetic symmetry. Amplitudes (3) include anapoles, t, and parity-even quadrupoles and octupoles $q = 2\sqrt{3} \langle \mathcal{T}^2_{+2} \rangle''$, $p = (3/2)\sqrt{7} \langle \mathcal{T}^3_0 \rangle$ and $p' = (3/2)\sqrt{(70/3)} \langle \mathcal{T}^3_{+2} \rangle'$. Multipoles in question are actually denoted $\langle T^K_Q \rangle$ in references [21, 22] but here we choose to reserve this particular notation for parity-even multipoles in x-ray diffraction encountered in Section IV. Moreover, a departure in notation is warranted on the grounds that we make explicit radial integrals $\langle j_n \rangle$ which in previous work are factors in multipoles. Observe that parity-even quadrupoles and anapoles do not contribute to $(\mathbf{\kappa} \cdot \mathbf{Q})$.

Radial integrals $\langle j_n \rangle$ with n = 0, 2, 4 & 6 calculated for $Ru^{4+}$ are shown in Fig. 2a (atomic code due to R. D. Cowan [27]). Evidently, $\langle j_4 \rangle$ and $\langle j_6 \rangle$ provide small corrections to an amplitude based on $\langle j_0 \rangle$ and $\langle j_2 \rangle$ with w < 6, and they are omitted from our analysis. This simplification directly affects octupoles, which are proportional to a linear combination of $\langle j_2 \rangle$ and $\langle j_4 \rangle$. Fig. 2b compares $\langle j_0 \rangle$, $\langle j_2 \rangle$ and $\langle j_4 \rangle$ for atomic configurations $4d^4$ and $4d^7$. There are significant differences for the two configurations, particularly with $\langle j_0 \rangle$, that influence data analysis. Also included in Fig. 2b are results for $Ru^{1+}$ ($4d^7$) derived from standard tabulations prepared by P. J. Brown [28] that are satisfyingly almost indistinguishable from our results for the same atomic configuration.

In fits of (2) and (3) to observed intensities we used $\langle L_\zeta \rangle = + 0.11$, and $\mu_\zeta = 1.8 \, \mu_B$. These results are consistent with an independent observation $|\langle L_\zeta \rangle|/\langle S_\zeta \rangle \approx 0.13$ [29]. Spin and orbital magnetic moments are parallel in the configuration $(t_{2g})^4$ appropriate for $Ru^{4+}$ in a strong octahedral ligand field [25]. As regard to the magnetic moment, Crawford *et al.*, [30] quote $\mu = 1.8(3) \, \mu_B$ for $Ru^{4+}$ in $Sr_4Ru_3O_6$, while Yoshida *et al.* [1] and Bao *et al.* [10] quote $\mu_\zeta = 1.59$

± .07 $\mu_B$ and $\mu_\zeta$ = 1.8(2) $\mu_B$, respectively, for $Ru^{4+}$ in $Ca_3Ru_2O_7$. Fits of $\Im(t)$ to observed intensities imply q ≈ 1.31, p ≈ 3.10, and p′ ≈ − 3.85, and these values are used in calculated values reported in Table I. Two values of t are obtained for most Bragg spots, because intensity (2) is a product of amplitudes that contain t. Table I lists experimental data [10], $\Im(t)$, t and $\Im(0)$ in ascending order of w. A goodness-of-fit $R_F$ = 1.2% with inclusion of anapoles slumps to 10 times this value when anapoles are omitted from calculated intensities.

Our results for t in Table I versus w are displayed in Fig. 3. In light of the excellent agreement between the observed cross-sections and $\Im(t)$ we attribute the evident scatter to limitations in the data that might be reduced on revisiting the experiment. A linear combination of radial integrals associated with spin, ($h_1$), and orbital, ($j_0$), anapoles is fitted to 12 selected values of t. The fit reported in Fig. 3 yields an estimate $\langle\Omega_\xi\rangle_S /\langle\Omega_\xi\rangle_L ≈ − (1/12)$.

The quality of our fit of $\Im(t)$ to intensities of Bragg spots $R_F$ = 1.2% is also achieved using oppositely aligned spin and orbital magnetic moments. Specifically, $\langle L_\zeta\rangle$ = − 0.125 and $\mu_\zeta$ = 1.8 $\mu_B$ imply q ≈ 1.17, p ≈ 2.90, and p′ ≈ − 3.50, and these multipoles yield an identical $R_F$ with changes to t that are entirely negligible.

## IV. RESONANT X-RAY DIFFRACTION

Bragg diffraction of x-rays tuned to a ruthenium atomic resonance is considered in this section [21, 31]. First, a study of published data [11], followed by feasibility studies for future experiments suggested by results from the foregoing analysis of neutron diffraction data. Unit-cell structure factors are derived from the electronic structure factor (1) that respects all elements of symmetry in the magnetic space-group $P_Cna2_1$. Results in Section IV. C incorporate a magnetic field applied along the b-axis with ferrimagnetic order described by the space group Pm′c′$2_1$. Throughout, we use universal expressions for unit-cell structure factors applicable to an azimuthal-angle scan, in which the crystal is rotated about the Bragg wavevector [23, 32]. States of polarization are defined in Fig. 4, and definitions of parity-even $\langle T^K_Q\rangle$ and Dirac $\langle G^K_Q\rangle$ multipoles comply with references [21, 23, 32, 33].

### A. Analysis of published data

An azimuthal-angle scan of intensity at the ($H_o$, $H_o$, 0) Bragg spot with $H_o$ = 1 and sample temperature 17 K is reported by Bohnenbuck *et al*. [11]. There is no intensity in the unrotated channel of polarization σ′σ with enhancement by an E1-E1 absorption event, because Bragg reflections of interest forbid charge-like (time-even) contributions. Magnetic contributions occur in both π′π and the two rotated channels. Data displayed in Fig. 5 are for the rotated channel of polarization σ′π.

With Miller indices (*h* + *l*) = 0 and phase angle φ = 0, the electronic structure factor (1) reduces to,

$$\Psi^K{}_Q = [1 + (-1)^Q] [\langle U^K{}_Q\rangle - \sigma_\pi (-1)^K \langle U^K{}_{-Q}\rangle]. \qquad (4)$$

Evidently, $\Psi^K{}_Q$ can be different from zero for projections Q even. Magnetic Bragg spots are indexed by $(h + k)$ odd. Recall that we use general coordinates $x = z = 1/4$, which is an approximation albeit a very good one [1]. In the subsequent calculation we add the further assumption that the Bragg wavevector (1, 1, 0) subtends 45º with the a-axis and the b-axis, which neglects a small difference between the cell lengths a and b (the angle between the a-axis and (1, 1, 0) is ≈ 44.12º). The two approximations mentioned do not modify our principal findings that rely on magnetic symmetry.

Magnetic diffraction enhanced by an E1-E1 event ($\sigma_\pi = +1$) is determined by a dipole $\langle T^1{}_Q\rangle$ and only $Q = 0$ is allowed by the electronic structure factor (4). The corresponding unit-cell structure factor for diffraction by Ru ions is found to be,

$$F^{(+)}{}_{\sigma'\pi}(1, 1) = - (i\langle T^1\zeta\rangle/2) [\cos(\theta) \cos(\psi) + \sin(\theta)]. \qquad (5)$$

We note in passing that $F^{(+)}{}_{\pi'\pi}(1, 1) = \{(i\langle T^1\zeta\rangle/2) \sin(2\theta) \sin(\psi)\}$.

The origin of the azimuthal angle ($\psi = 0$) in (3) is such that a- and b-crystal axes span the plane of scattering [11]. The Bragg angle $\theta \approx 32.8º$ for (1, 1, 0) and an x-ray energy 2.965 keV for the $L_2$ absorption edge of $Ru^{4+}$ ($\sin(\theta) \approx 0.130$ ($\lambda H_o$) Å$^{-1}$ with $\lambda \approx 12.40/E$ and photon wavelength $\lambda$ and primary energy E in units of Å and keV, respectively). Note that the dipole $\langle T^1\zeta\rangle$ is purely real while $F^{(+)}{}_{\sigma'\pi}(1, 1)$ is purely imaginary. Intensity in the $\sigma'\pi$ channel is proportional to $I(1, 1) = |F^{(+)}{}_{\sigma'\pi}(1, 1)|^2$, and it is confronted with available experimental data in Fig. 5a. The proportionality factor for $I(1, 1)$ contains a dipole radial integrals $\langle 2p|r|4d\rangle$ or $\langle 3p|r|4d\rangle$ about which we say more later in the context of relative strengths of E1-E1, E1-M1 and E1-E2 diffraction amplitudes.

In the E1-E2 parity-odd event, Dirac multipoles $\langle G^K{}_Q\rangle$ have ranks $K = 1, 2$ and $3$. Anapoles ($K = 1$) are forbidden by (4), however, and three multipoles contribute to the diffraction pattern. The unit-cell structure factor can be neatly expressed in terms of two functions of $\theta$ and $\psi$, namely,

$$\nu = [\sin^2(\theta) - \{\cos(\theta) \cos(\psi)\}^2] \text{ and } \chi = [2 \sin(2\theta) \cos(\psi)], \qquad (6)$$

and we find,

$$F^{(-)}{}_{\sigma'\pi}(1, 2) = (1/4\sqrt{5}) \Re(1, 2) [\langle G^2{}_0\rangle \{\nu + \chi\} + \sqrt{(2/3)} \langle G^2{}_{+2}\rangle' \{3\nu - \chi\} \qquad (7)$$

$$+ (2/\sqrt{3}) \langle G^3{}_{+2}\rangle'' \chi \{3 \cos^2(\psi) - 2\}].$$

Here, $\langle G^2_{+2}\rangle'$ and $\langle G^3_{+2}\rangle''$ are real and imaginary parts of a quadrupole and an octupole, respectively. The parity-even and parity-odd structure factors (5) and (7) are seen to differ by a 90° phase. While the azimuthal-angle dependence of $F^{(+)}_{\sigma'\pi}(1, 1)$ is simply one harmonic, $\cos(\psi)$, $F^{(-)}_{\sigma'\pi}(1, 2)$ also includes harmonics $\cos(2\psi)$ and $\cos(3\psi)$ as a consequence of quadrupoles and an octupole. Intensities $I(1, 1)$ and $I(1, 2) = |F^{(-)}_{\sigma'\pi}(1, 2)|^2$ are symmetric around $\psi = 180°$.

Data in Fig. 5a [11] are in conflict with our structure factors because they are not symmetric about $\psi = 180°$. However, intensities $I(1, 1)$ and $I(1, 2)$, derived from (5) and (7), respectively, fit the data very well if the origin of the azimuthal-angle scan is offset, which is reported in Figs. 5b, c. For the moment, offsets in the azimuthal angle that create an accord between data and calculated intensities are attributed to experimental conditions (we thank Dr E. Schierle for clarification [11]). Revisiting resonant x-ray Bragg diffraction might add insight to the problem of asymmetry in diffraction at $(H_o, H_o, 0)$. Feasibility studies of different Bragg spots are reported in the next sub-section with this problem in mind. The octupole in $I(1, 2)$ is a mismatch to experimental data and only the allowed Dirac quadrupoles are used in the fit shown in Fig. 5c.

To be meaningful, structure factors (5) and (7) should include strengths of E1-E1 and E1-E2 events, in particular radial integrals of the position variable taken between the core state and valence states. In our case, the ratio of E1-E2 to E1-E1 strengths is expressed by a dimensionless factor included in the result (7) for $F^{(-)}_{\sigma'\pi}(1, 2)$. With the photon energy E tuned to an L-edge, 2p,

$$\Re(1, 2) = [(\alpha E)/(2\, a_o\, R_\infty)]\, (\langle 2p|r^2|5p\rangle/\langle 2p|r|4d\rangle), \qquad (8)$$

where $\alpha$, $a_o$ and $R_\infty$ are the fine structure constant, Bohr radius and the Rydberg unit of energy, respectively. A relativistic atomic code provides the estimates $(\langle 2p|r^2|5p\rangle/\langle 2p|r|4d\rangle) = -0.047\, a_o$ while $(\langle 2p|r^2|4f\rangle/\langle 2p|r|4d\rangle) = -0.031\, a_o$ [27].

B. Chiral order

We report unit-cell structure factors for magnetic $(0, 0, L_o)$ Bragg spots, with odd $L_o$. Our analysis of the corresponding Bragg spots in neutron diffraction patterns, Section III, indicates that anapoles are significant, and it is reasonable to conjecture that higher-order Dirac multipoles are likewise significant. The conjecture is supported by results from the simulation of electronic structure performed by Thöle and Spaldin [34]. These authors include the monopole in their study and it contributes to diffraction enhanced by an E1-M1 absorption event, and witnessed in the structure factor (12). We continue with calculations of structure factors for E1-E1, E1-M1 and E1-E2 events.

The unit-cell structure factor for E1-E1 and $(0, 0, L_o)$ is,

$$F^{(+)}{}_{\sigma'\pi}(1, 1) = (1/\sqrt{2}) [i\langle T^1_\zeta\rangle \cos(\varphi) \cos(\theta) \sin(\psi) - \langle T^1_\eta\rangle \sin(\varphi) \sin(\theta)], \qquad (9)$$

with $F^{(+)}{}_{\pi'\sigma}(1, 1) = \{F^{(+)}{}_{\sigma'\pi}(1, 1)\}^*$. Recall that $\varphi = (2\pi y L_o)$ with $\sin(\varphi) \approx 0$ for $L_o = 5$, to a good approximation. The b-axis is normal to the plane of scattering for $\psi = 0$. Note in (9) that the contribution using the secondary dipole $\langle T^1_\eta\rangle = \langle T^1_c\rangle$ is independent of the azimuthal angle, as it should be when the Bragg and propagation (c-axis) vectors are collinear. Intensity derived from (9) has a $\sin^2(\psi)$ dependence. The Bragg angle satisfies $\sin(\theta) \approx 0.316 (L_o/E)$ with E in units of keV. Thus, only $L_o = 1$ is accessible at M-edges (3p) and $E \approx 0.47$ keV, while $L_o = 1$, 3, 5, 7 are accessible at L-edges and $E \approx 3.0$ keV.

The 90° phase difference between the two axial dipoles in (9) indicates a chiral (handed) order that manifests itself with intensity that depends on circular polarization in the primary x-ray beam. Intensity in the handed setting is proportional to the imaginary part of interference between rotated and unrotated channels of polarization [35]. Since $F^{(+)}{}_{\sigma'\sigma}(1, 1) = 0$ for magnetic reflections $(0, 0, L_o)$ the E1-E1 intensity of interest reduces to,

$$\Pi^{(+)} = P_2 \, \text{Im}.\{[F^{(+)}{}_{\pi'\pi}(1, 1)]^* \, F^{(+)}{}_{\pi'\sigma}(1, 1)\}$$

$$= -(P_2/2) \sin(2\varphi) \sin^2(\theta) \cos(\theta) \cos(\psi) \langle T^1_\zeta\rangle \langle T^1_\eta\rangle. \qquad (10)$$

Here, $P_2$ is the Stokes parameter for primary circular polarization [21, 35]. Evidently, intensity in the proposed experiment is an interference between the primary and secondary axial dipoles.

Our first parity-odd event is E1-M1. Dipoles M1 and E1 in $\Re(1, 1)$, the analogue of (8), have magnitudes $\mu_B$ and $(e\, a_o)$, respectively, where $\mu_B$ is the Bohr magneton. Using $\mu_B/(e\, a_o) = \alpha/2$ leads to,

$$\Re(1, 1) = (\alpha\, a_o \langle \lambda|\lambda'\rangle)/\langle 2p|r|4d\rangle. \qquad (11)$$

In this expression, $\langle \lambda|\lambda'\rangle$ is the overlap of orbitals in the M1 event that possess the same angular momentum, because the magnetic moment operator is diagonal in this basis [36]. We go on to find,

$$F^{(-)}{}_{\sigma'\pi}(1, 1) = -(i/\sqrt{2}) \Re(1, 1) \langle G^1_\xi\rangle \sin(\varphi) \sin(2\theta) \sin(\psi)$$

$$- (2/\sqrt{3}) \Re(1, 1) \cos(\varphi) [\langle G^0_0\rangle \sin^2(\theta) + (1/2\sqrt{2}) \langle G^2_0\rangle \{1 + \cos^2(\theta) [2 - 3 \sin^2(\psi)]\}$$

$$+ (\sqrt{3}/2) \langle G^2_{+2}\rangle'\{1 + [\cos(\theta) \sin(\psi)]^2\}]. \qquad (12)$$

and $F^{(-)}{}_{\pi'\sigma}(1, 1) = \{F^{(-)}{}_{\sigma'\pi}(1, 1)\}^*$. The anapole $\langle G^1_\xi\rangle$ is depicted in Fig. 1, and its contribution in (12) is out of phase with the contribution from even-rank multipoles. The magnetic monopole, $\langle G^0_0\rangle$, carries no dependence on the azimuthal angle, of course [37]. Comparing (9) and (12) we see that, dipoles $\langle T^1\zeta\rangle$ and $\langle G^1\xi\rangle$ are accompanied by $\sin(\psi)$, with the anapole contribution in (12) very small for $L_o = 5$.

Dirac multipoles possess a chiral configuration. It is exposed in a handed setting provided by circular polarization in the primary beam of x-rays, and parallel Bragg and propagation vectors. Structure factors for all four channels of polarization contribute to the E1-M1 intensity in question, unlike the corresponding calculation for E1-E1 that is significantly simplified by the result $F^{(+)}_{\sigma'\sigma}(1, 1) = 0$. In the present case, E1-M1 intensity is given by [35],

$$\Pi^{(-)} = - (4 P_2/\sqrt{3}) \, \mathfrak{R}^2(1, 1) \sin(2\varphi) \cos(\theta) \cos(\psi) \langle G^2_{+1}\rangle''$$

$$\times [\langle G^0_0\rangle \sin^2(\theta) + (1/2\sqrt{2}) \langle G^2_0\rangle \{1 + \cos^2(\theta) [2 - 3 \sin^2(\psi)]\}$$

$$+ (\sqrt{3}/2) \langle G^2_{+2}\rangle' \{1 + [\cos(\theta) \sin(\psi)]^2\}]. \qquad (13)$$

Notably, intensity does not depend on the anapole, whereas corresponding intensity $\Pi^{(+)}$ created in an E1-E1 event arises from dipoles alone. However, (10) and (13) are both proportional to $\sin(2\varphi) \cos(\psi)$.

Octupoles in the E1-E2 structure factor are omitted in subsequent calculations, principally on the grounds that they are at odds with data for (1, 1, 0) but also because the dominant contribution to the diffraction amplitude is likely given by an anapole $\langle G^1\xi\rangle$. Including all dipoles and quadrupoles allowed in $(0, 0, L_o)$ reflections we find,

$$F^{(-)}_{\sigma'\pi}(1, 2) \approx \mathfrak{R}(1, 2) \, \{(2i/5) \sqrt{3} \, [\{\langle G^1\xi\rangle - (2/3)\sqrt{10} \langle G^2_{+1}\rangle''\} \sin(\varphi) \sin(2\theta) \sin(\psi)]$$

$$- (1/2) \sqrt{(1/5)} \cos(\varphi) [\langle G^2_0\rangle \nu + \sqrt{(2/3)} \langle G^2_{+2}\rangle' \{6 \sin^2(\theta) - \nu - 2\}]\}, \qquad (14)$$

with $\nu$ defined in (6), and $F^{(-)}_{\pi'\sigma}(1, 2) = \{F^{(-)}_{\sigma'\pi}(1, 2)\}^*$. The two components of (14), which differ by a 90° phase, can be separated in an experiment by comparing data for different Miller indices using $\sin(\varphi) \approx 0$ for $L_o = 5$.

### C. Applied magnet field

A magnetic field applied to $Ca_3Ru_2O_7$ induces a (first-order) metamagnetic transition to a canted antiferromagnetic structure in which axial dipoles are partially polarized in the direction of the field. In the low temperature phase, T < 48 K, a canted structure occurs in a field that exceeds ≈ 5.5 T parallel to the b-axis [10, 13]. In such a magnetic field, a small magneto-resistive effect occurs in the transition [10]. The metamagnetic transition no longer occurs at temperatures above the metal-to-nonmetal transition at T = 48 K.

The precursor to the transition is a ferrimagnetic order described by the space group Pm′c′2₁ (#26.70) [20]. The crystal class m′m′2 is asymmetric, polar and compatible with ferromagnetism. Bulk magnetization parallel to the applied field (b-axis) comes from two non-equivalent Ru sites that are allowed to have different axial magnetic dipoles moments in the zero-field A-type antiferromagnetic order described by $P_Cna2_1$.

A basis (0, 0, 1), (1, 0, 0), (0, 1, 0) with respect to the $Bb2_1m$ paramagnetic space-group defines principal axes $(\xi, \eta, \zeta) \equiv$ (c, a, b) not shown in Fig. 1, with Miller indices $h = L_o$, $k = H_o$, $l = K_o$. The applied field is parallel to the $\zeta$-axis. The electronic structure factor for $Pm'c'2_1$ is,

$$\Psi^K_Q = \exp(i2\pi zl) \{\langle U^K_Q \rangle [\exp(i\varphi) + (-1)^{l+Q} \exp(-i\varphi)] \quad (15)$$

$$+ \sigma_\theta \sigma_\pi (-1)^{l+K+Q} \langle U^K_{-Q} \rangle [\exp(i\varphi') + (-1)^{l+Q} \exp(-i\varphi')]\},$$

with $\varphi = 2\pi(xh + yk)$ and $\varphi' = 2\pi(xh - yk)$, and general coordinates (x, y, z). Bulk magnetization is revealed by a non-zero value of $\Psi^K_Q(Pm'c'2_1)$ evaluated for $h = k = l = 0$, $K = 1$ and $\sigma_\theta \sigma_\pi = -1$ (axial magnetism). For these conditions, the structure factor (15) is non-zero for Q = 0, i.e., a bulk ferromagnetic moment parallel to the b-axis.

Restrictions for purely magnetic Bragg spots are identified in (15) evaluated with Q = 0, K even and $\sigma_\theta \sigma_\pi = +1$ appropriate to charge or nuclear scattering. With said conditions, $\Psi^K_0 = 0$ for (h, 0, l), l odd and no restrictions on h. (In the absence of a magnetic field, using space group $P_Cna2_1$, the corresponding restrictions are $(h + l) = (H_o + K_o)$ even with $(h + k) = (H_o + L_o)$ odd arising from anti-translation.) Unit-cell structure factors for purely magnetic diffraction are derived from,

$$\Psi^K_Q = \exp(i2\pi zl) [\exp(i\varphi) - (-1)^Q \exp(-i\varphi)] \quad (16)$$

$$\times [\langle U^K_Q \rangle - \sigma_\theta \sigma_\pi (-1)^{K+Q} \langle U^K_{-Q} \rangle],$$

which is obtained by setting $k = 0$ and $l$ odd in (15). Spatial phases are reduced to $\varphi = \varphi' = (2\pi xh) = (2\pi xL_o)$, while $(h, 0, l) \equiv (0, K_o, L_o)$ with $K_o$ odd and all $L_o$. The leading phase factor in (16) can be set aside in the calculation of Bragg intensities.

Ruthenium ions occupy non-equivalent sites 4c that possess no symmetry. The two sites have general coordinates (x, y, z) ≈ (0.4008, 1/4, 3/4) and (0.4008 − 1/2, 3/4, 3/4). For purely magnetic Bragg spots $\varphi = (2\pi xL_o)$ with x ≈ 0.4008 [1], and $\varphi$ is identical to the spatial angle used in all foregoing calculations; phase factors for the two Ru sites are $\exp(i\varphi)$ and $(-1)^{L_o} \exp(i\varphi)$.

With a photon energy E ≈ 3.0 keV for Ru L-edges and $K_o = 1$ the Bragg condition can be met with $L_o = 0, 1, 2, ... , 7$, while $K_o > 1$ is not achievable with the specified energy. The E1-E1 unit-cell structure factor in the rotated channel of polarization is,

$$F^{(+)}_{\sigma'\pi}(1, 1) = - (i\langle T^1_\eta\rangle/2) \cos(\varphi) \cos(\theta) \sin(\psi). \qquad (17)$$

while $F^{(+)}_{\sigma'\sigma}(1, 1) = 0$. The axial dipole $\langle T^1_\eta\rangle = \langle T^1_a\rangle$ is normal to the plane of scattering when the azimuthal angle $\psi = 0$. It is notable that $F^{(+)}_{\sigma'\pi}(1, 1)$, as well as $F^{(+)}_{\pi'\pi}(1, 1)$, does not specifically depend on the orientation of magnetic field, which is along the b-axis. Also, there is no chiral order in axial dipoles to be exposed by circular polarization and $\Pi^{(+)} = 0$.

Parity-odd diffraction enhanced by an E1-M1 event does not contain a monopole. Instead, diffraction engages anapoles parallel ($\langle G^1_\zeta\rangle$) and perpendicular ($\langle G^1_\xi\rangle$) to the magnetic field, and two quadrupoles. For the rotated channel of polarization,

$$F^{(-)}_{\sigma'\pi}(1, 1) = 2\,\mathfrak{R}(1, 1) \cos(\theta) \sin(\psi) \{\cos(\varphi) [- (1/\sqrt{2}) \langle G^1_\xi\rangle \sin(\alpha) \sin(\theta) + A\,\langle G^2_{+1}\rangle'']$$

$$- i \sin(\varphi) [(1/\sqrt{2}) \langle G^1_\zeta\rangle \cos(\alpha) \sin(\theta) + B\,\langle G^2_{+2}\rangle'']\}, \qquad (18)$$

with factors $A = [\cos(\alpha) \cos(\theta) \cos(\psi)]$ and $B = [\sin(\alpha) \cos(\theta) \cos(\psi)]$. The angle $\alpha$ arises from the orientation of the b-axis (field direction) within the plane of scattering at $\psi = 0$, and it finds no place in E1-E1 structure factors. Specifically,

$$\cos(\alpha) = - L_o/\sqrt{[(cK_o/b)^2 + (L_o)^2]}, \qquad (19)$$

with $\alpha = \pi/2$ for $L_o = 0$, which also means $\varphi = 0$. The Bragg wavevector $(0, K_o, 0)$ is parallel to the magnetic field and $F^{(-)}_{\sigma'\pi}(1, 1) \propto [\langle G^1_\xi\rangle \sin(2\theta) \sin(\psi)]$, because $\sin(\varphi) = 0$ and $A = 0$. In the general case, $F^{(-)}_{\sigma'\pi}(1, 1)$ contains two harmonics of the azimuthal angle, namely, $\sin(\psi)$ and $\sin(2\psi)$.

Diffraction in the rotated channel of polarization enhanced by an E1-E2 absorption event is described by a structure factor similar to (18) when octupoles are neglected as they are in (14). In this approximation, $F^{(-)}_{\sigma'\pi}(1, 1)$ and $F^{(-)}_{\sigma'\pi}(1, 2)$ have identical factors from the azimuthal angle. Indeed, $F^{(-)}_{\sigma'\pi}(1, 2)$ can be derived from (18) with the substitutions $A \to (1/3)\sqrt{5}\,[A - 2 \sin(\alpha) \sin(\theta)]$, $B \to (1/3)\sqrt{5}\,[B + 2 \cos(\alpha) \sin(\theta)]$ and $\mathfrak{R}(1, 1) \to \mathfrak{R}(1, 2)$, apart from unimportant numerical factors. The change to A means that $F^{(-)}_{\sigma'\pi}(1, 2)$ depends on two multipoles, $\langle G^1_\xi\rangle$ and $\langle G^2_{+1}\rangle''$, for $(0, K_o, 0)$.

## V. CONCLUSIONS

In our study of antiferromagnetic $Ca_3Ru_2O_7$ we have presented:
- the magnetic space-group $P_Cna2_1$ (crystal class mm21′) that is appropriate for A-type antiferromagnetic order
- compelling evidence derived from a neutron diffraction pattern [10] for the existence of Dirac dipoles (anapoles)

- radial integrals in neutron diffraction amplitudes for $Ru^{4+}$ ($4d^4$)
- inconclusive evidence for Dirac multipoles derived from published diffraction pattern obtained with x-rays tuned to a ruthenium atomic resonance [11]
- chiral order of both axial dipoles, created by primary and secondary dipoles, and Dirac multipoles that can be exposed in the resonant diffraction of circularly polarized x-rays
- the magnetic space-group Pm′c′2$_1$ (crystal class m′m′2) that is appropriate for ferrimagnetic order induced by a magnetic field applied parallel to the crystal b-axis, together with feasibility studies of pertinent resonant x-ray diffraction experiments.

Our evidence for anapoles in the low-temperature magnetic configuration is bolstered by their appearance in a simulation of the electronic structure [34].

**Acknowledgements.** We are grateful to Dr S. Agrestini, Dr A. Bombardi, Professor J. W. Lynn, Professor A. Michels, Dr E. Schierle and Dr D. A. Sokolov for useful advice and opinions.

TABLE I. Observed neutron cross-sections $\sigma_{obs}$ listed against increasing w. Sample temperature = 3.5 K. Bragg spots are here labelled by orthorhombic Miller indices $(H_o, K_o, L_o)$. Inferred values of anapoles, t, using $\langle L\zeta \rangle = +0.11$ and $\mu\zeta = \langle 2S\zeta + L\zeta \rangle = 1.80$. Calculated values of the cross-section $\Im(t)$ derived from (2) and (3) in the fourth column possess a goodness-of-fit $R_F = 1.2\%$, with $R_F$ defined in standard manner $R_F = \{\sum |\sigma_{obs} - \Im(t)| / \sum \sigma_{obs}\}$ with both sums over all 15 Bragg spots. Omission of anapoles in $\Im(0)$ yields a vastly inferior $R_F = 13.6\%$. For $L_o = 5$ no realistic value can be assigned to t because the corresponding structural phase factor $\sin(\varphi) \approx 0.0$.

| $(H_o, K_o, L_o)$ | w | $\sigma_{obs}$ | $\Im(t)$ | t | $\Im(0)$ |
|---|---|---|---|---|---|
| (0, 0, 1) | 0.51 | 9.64(3) | 9.64 | −0.01 | 9.57 |
| (0, 0, 3) | 1.53 | 1.10(2) | 1.09 | +0.035, −0.98 | 0.95 |
| (0, 0, 5) | 2.56 | 5.91(6) | 5.91 | ~ | |
| (0. 0, 7) | 3.58 | 0.43(2) | 0.43 | +0.56, −0.09 | 0.23 |
| (0, 2, 1) | 3.65 | 0.070(7) | 0.070 | +0.39, −0.04 | 0.05 |
| (2, 0, 1) | 3.75 | 2.39(1) | 2.41 | +0.50 | 2.69 |
| (0, 2, 3) | 3.92 | 0.051(7) | 0.047 | +0.01, −0.20 | 0.038 |
| (2, 0, 3) | 4.02 | 0.216(5) | 0.217 | −0.05 | 0.26 |
| (0, 2, 5) | 4.43 | 0.44(2) | 0.44 | ~ | |
| (2, 0, 5) | 4.51 | 1.66(1) | 1.43 | ~ | |
| (0, 0, 9) | 4.60 | 1.34(5) | 1.33 | +0.60, −1.13 | 0.14 |
| (0, 2, 7) | 5.09 | 0.05(1) | 0.05 | +0.175, −0.05 | 0.017 |
| (0, 0, 11) | 5.63 | 0.63(4) | 0.63 | +0.47, −0.88 | 0.055 |
| (0, 2, 9) | 5.85 | 0.39(3) | 0.38 | +0.48, −0.44 | 0.0 |
| (0, 2, 11) | 6.69 | 0.27(3) | 0.27 | +0.14, −0.75 | 0.12 |

FIG. 1. Crystal and magnetic structures of $Ca_3Ru_2O_7$ (cell lengths a ≈ 5.3677, b ≈ 5.5356, c ≈ 19.5219 Å at 8 K). Anapoles parallel to the crystal a-axis are depicted together with principal axial dipoles parallel to the crystal b-axis. Axes $(\xi, \eta, \zeta) \equiv (a, c, -b)$ for the magnetic structure $P_Cna2_1$ are displayed.

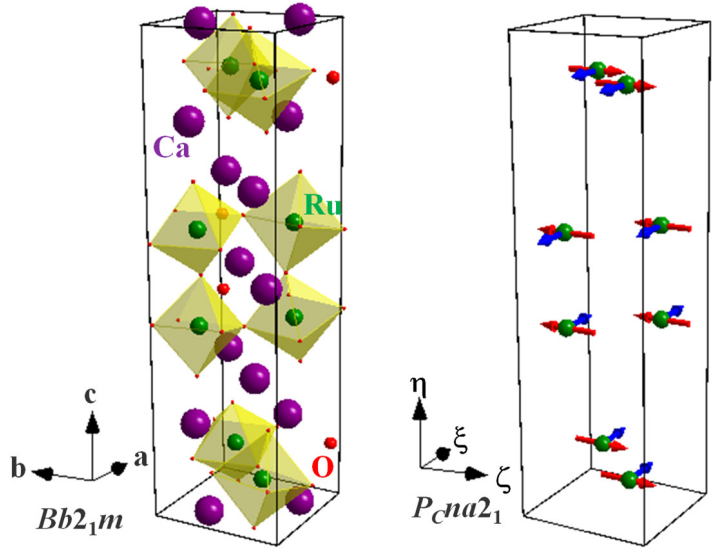

FIG. 2. (a) Radial integrals for Dirac multipoles that appear in the anapole $t = [3\langle\Omega\xi\rangle_S (h_1) - \langle\Omega\xi\rangle_L (j_0)]$ using a dimensionless variable $w = 12\pi a_o s$, where $a_o$ is the Bohr radius and $s = \sin(\theta)/\lambda$. Atomic wavefunctions $4d^4 - 5p^1$ (see text). Legend: (—) $(h_1)$, (—) $[w \times (j_0)]$ and (—) $[w \times (g_1)/10]$. Note that $(j_0)$ and $(g_1)$ are proportional to $1/w$ as the wavevector approaches zero, and $t' = \{\langle n\xi\rangle (g_1)\}$. Radial integrals $\langle j_0\rangle$ (black), $\langle j_2\rangle$ (orange), $\langle j_4\rangle$ (brown) and $\langle j_6\rangle$ (black) for axial multipoles using $Ru^{4+}$ ($4d^4$). (b) $\langle j_0\rangle$, $\langle j_2\rangle$ and $\langle j_4\rangle$ using black continuous curves for the configuration $4d^4$, and red continuous $4d^7$. Green dashed are results for $4d^7$ obtained from a standard tabulation (see text).

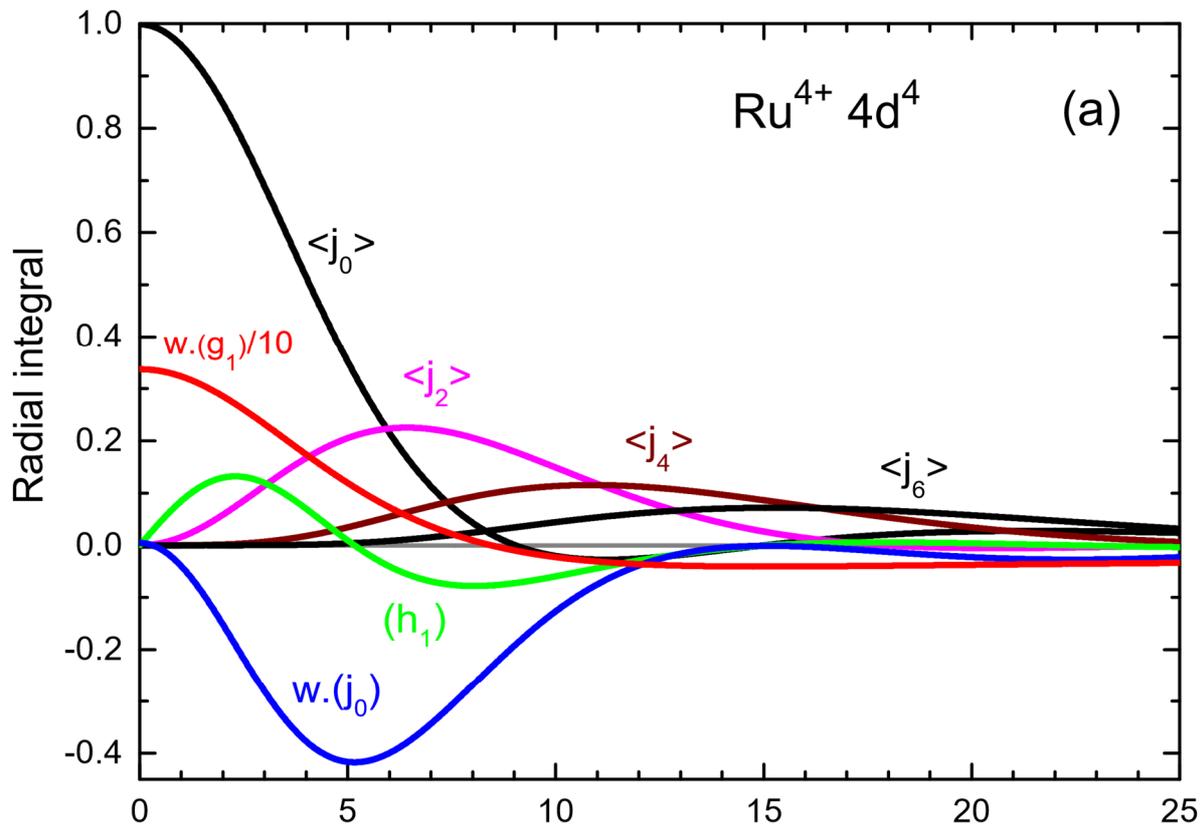
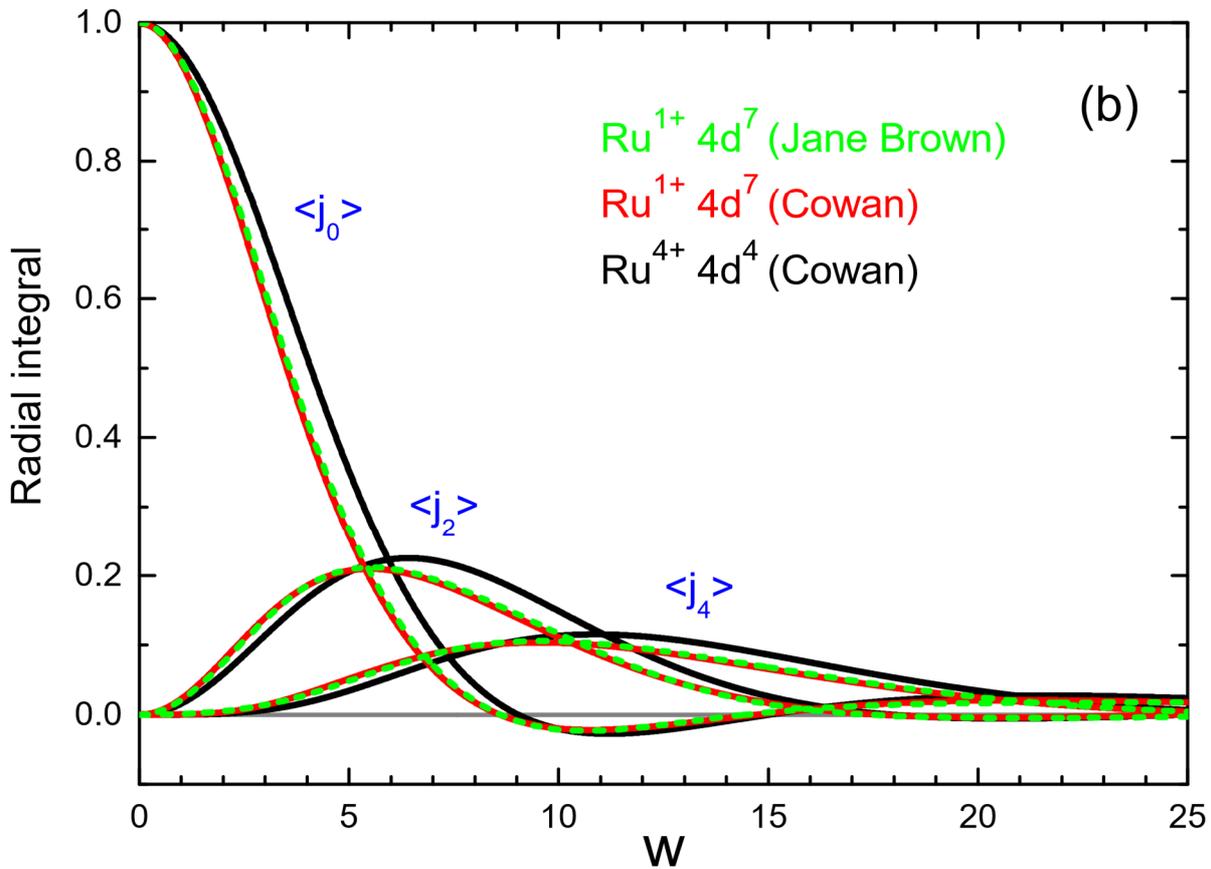

FIG. 3. Values of the anapole t listed in Table I are displayed as a function of w using black squares and red dots. For most Bragg spots there are two values of t, one black and one red, while no value can be assigned to t at Bragg spots ($H_o$, $K_o$, $L_o$) with $L_o = 5$. The solid curve is a fit to a linear combination of radial integrals for spin and orbital anapoles reported in Fig. 2a to 12 values of t denoted by black squares.

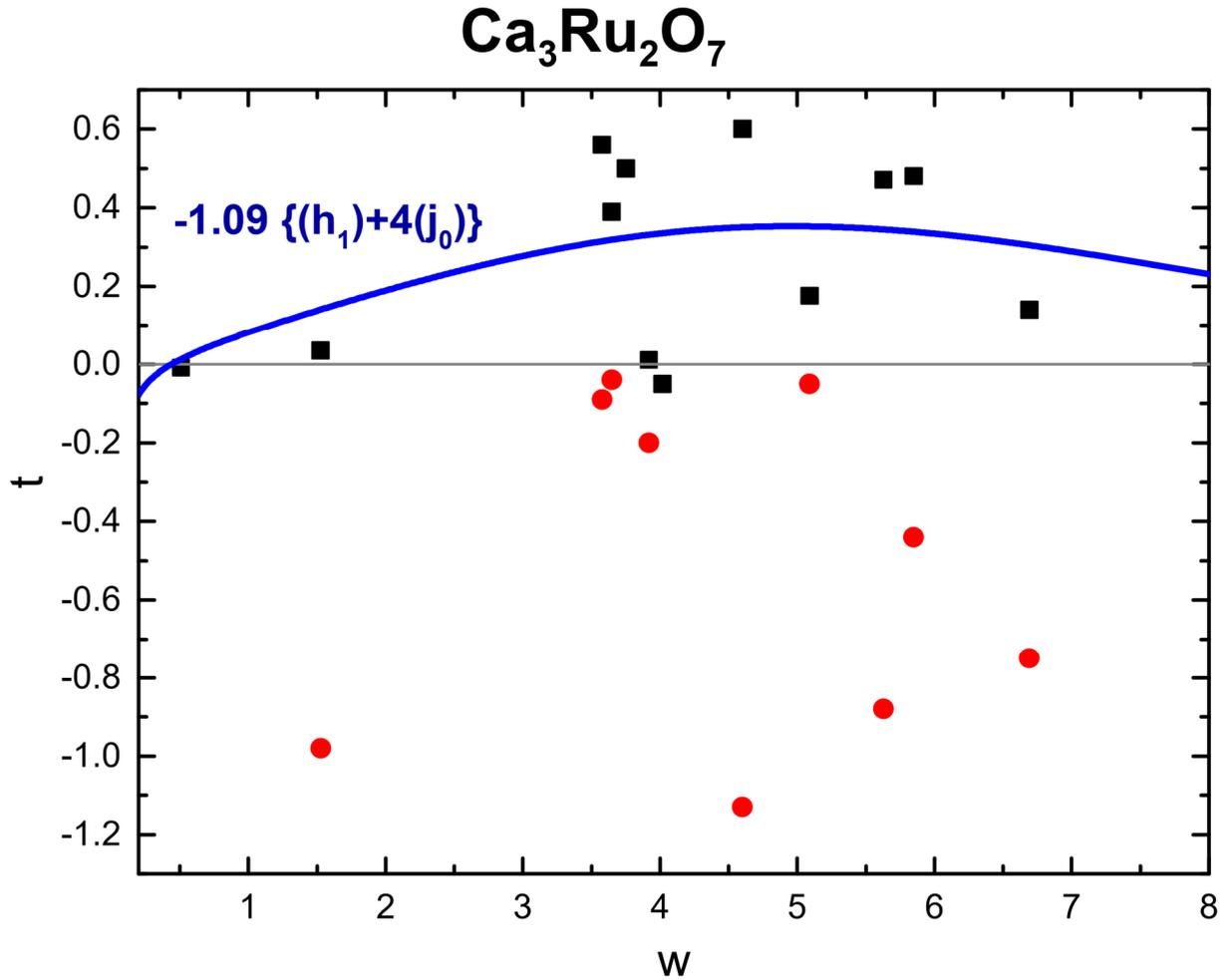

FIG. 4. Primary (σ, π) and secondary (σ′, π′) states of polarization. Corresponding wavevectors **q** and **q**′ subtend an angle 2θ, and **k** = **q** − **q**′. Principal axes (ξ, η, ζ) for a magnetic structure and depicted Cartesian co-ordinates (x, y, z) coincide in the nominal setting, and **k** for a particular Bragg spot is aligned with − **x**.

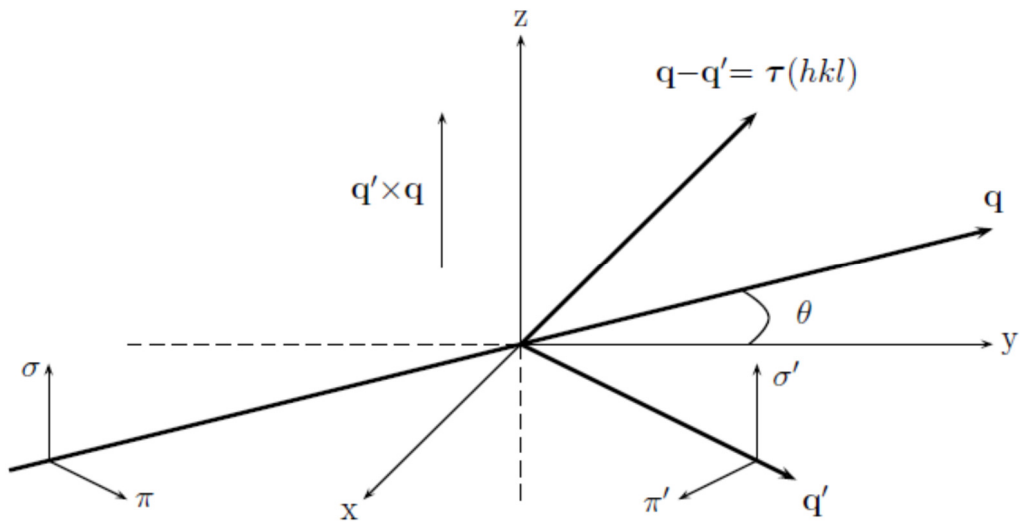

FIG. 5. Green triangles: Measured intensity of the (1, 1, 0) Bragg spot in the σ′π channel of polarization as a function of azimuthal angle, ψ. (a) Symmetric and antisymmetric components

of the data are displayed as red and blue spots, respectively. Continuous green curve is intensity I(1, 1) = $|F^{(+)}_{\sigma'\pi}(1, 1)|^2$, derived from (5) for enhancement by an E1-E1 event. Origin $\psi = 0$ at which a- and b-axes span the plane of scattering defined in Fig. 4. (b) Continuous red curve is intensity I(1, 1) using an $\psi$-origin off-set by 9.54º. (c) Continuous red curve is intensity I(1, 2) derived from (7) for an E1-E2 event. The $\psi$-origin is off-set by 12.39º and allowed Dirac octupoles are set equal to zero in the calculation.

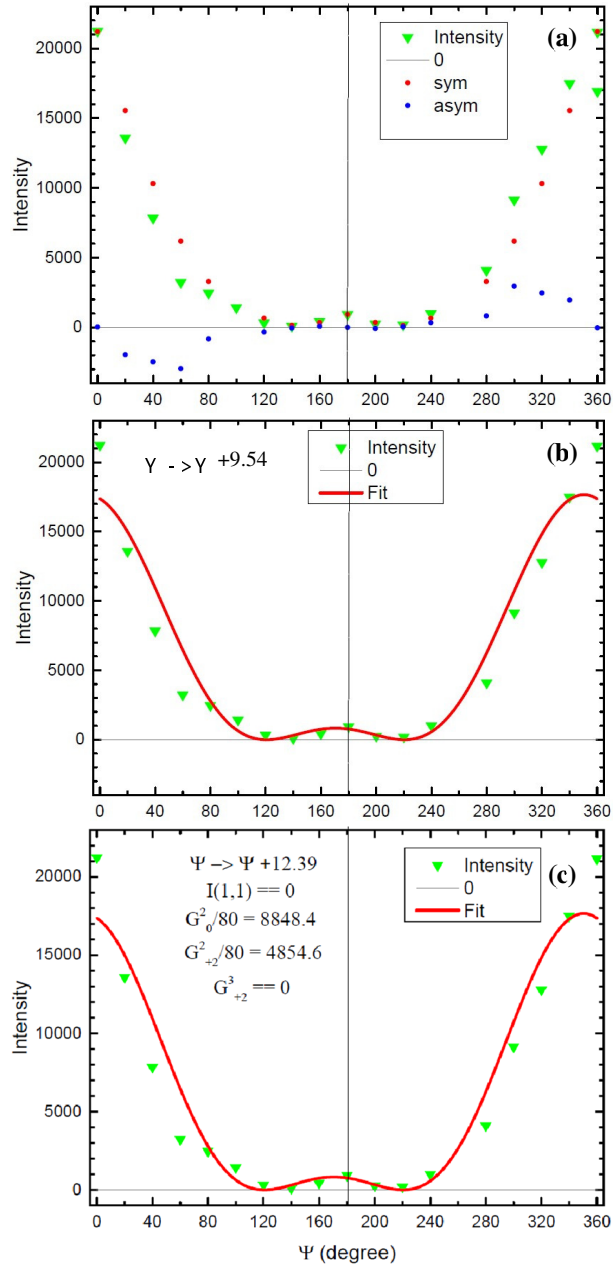